\documentclass[10pt,journal,compsoc]{IEEEtran}

\ifCLASSOPTIONcompsoc
  \usepackage[nocompress]{cite}
\else
  \usepackage{cite}
\fi

\ifCLASSOPTIONcompsoc
 \usepackage[caption=false,font=footnotesize,labelfont=sf,textfont=sf]{subfig}
\else
 \usepackage[caption=false,font=footnotesize]{subfig}
\fi

\ifCLASSINFOpdf
  \usepackage[pdftex]{graphicx}
  \graphicspath{{./graphics/}}
  \DeclareGraphicsExtensions{.pdf,.jpg,.png}
\else
  \usepackage[dvips]{graphicx}
  \graphicspath{{./graphics/}}
  \DeclareGraphicsExtensions{.eps}
\fi

\usepackage{amsmath}
\usepackage{algorithmic}
\usepackage{array}
\usepackage{stfloats}
\usepackage{url}
\usepackage{algorithm}
\usepackage{textcomp}
\usepackage{verbatim}
\usepackage{enumitem}
\usepackage{hhline}
\usepackage[table,xcdraw]{xcolor}
\usepackage{multirow}
\usepackage{footnote}

\begin{document}
\title{SAL-PIM: A Subarray-level Processing-in-Memory Architecture with LUT-based Linear Interpolation for Transformer-based Text Generation}

\author{Wontak Han,~\IEEEmembership{Student Member,~IEEE,} 
        Hyunjun Cho,~\IEEEmembership{Student Member,~IEEE,} 
        \\Donghyuk Kim,~\IEEEmembership{Student Member,~IEEE,} 
        and Joo-Young Kim,~\IEEEmembership{Senior Member,~IEEE} 
\IEEEcompsocitemizethanks{\IEEEcompsocthanksitem The authors are with the Department of Electrical Engineering, Korea Advanced Institute of Science and Technology (KAIST), Daejeon 34141, South Korea.\protect\\
E-mail: \{11tak, jooyoung1203\}@kaist.ac.kr}
\thanks{Manuscript received April 19, 2005; revised August 26, 2015.}}

\markboth{.}%
{Shell \MakeLowercase{\textit{et al.}}: Bare Advanced Demo of IEEEtran.cls for IEEE Computer Society Journals}

\IEEEtitleabstractindextext{%
\begin{abstract}

Text generation is a compelling sub-field of natural language processing, aiming to generate human-readable text from input words. Although many deep learning models have been proposed, the recent emergence of transformer-based large language models advances its academic research and industry development, showing remarkable qualitative results in text generation.
In particular, the decoder-only generative models, such as generative pre-trained transformer (GPT), are widely used for text generation, with two major computational stages: summarization and generation. Unlike the summarization stage, which can process the input tokens in parallel, the generation stage is difficult to accelerate due to its sequential generation of output tokens through iteration. Moreover, each iteration requires reading a whole model with little data reuse opportunity. Therefore, the workload of transformer-based text generation is severely memory-bound, making the external memory bandwidth system bottleneck.

In this paper, we propose a subarray-level processing-in-memory (PIM) architecture named SAL-PIM, the first HBM-based PIM architecture for the end-to-end acceleration of transformer-based text generation. With optimized data mapping schemes for different operations, SAL-PIM utilizes higher internal bandwidth by integrating multiple subarray-level arithmetic logic units (S-ALUs) next to memory subarrays. To minimize the area overhead for S-ALU, it uses shared MACs leveraging slow clock frequency of commands for the same bank. In addition, a few subarrays in the bank are used as look-up tables (LUTs) to handle non-linear functions in PIM, supporting multiple addressing to select sections for linear interpolation.
Lastly, the channel-level arithmetic logic unit (C-ALU) is added in the buffer die of HBM to perform the accumulation and reduce-sum operations of data across multiple banks, completing end-to-end inference on PIM. To validate the SAL-PIM architecture, we built a cycle-accurate simulator based on Ramulator. We also implemented the SAL-PIM's logic units in 28-nm CMOS technology and scaled the results to DRAM technology to verify its feasibility. 
We measured the end-to-end latency of SAL-PIM when it runs various text generation workloads on the GPT-2 medium model (with 345 million parameters), in which the input and output token numbers vary from 32 to 128 and from 1 to 256, respectively. As a result, with 4.81\% area overhead, SAL-PIM achieves up to 4.72$\times{}$ speedup (1.83$\times{}$ on average) over the Nvidia Titan RTX GPU running FasterTransformer Framework.

\end{abstract}

}

\maketitle

\IEEEraisesectionheading{\section{Introduction}\label{sec:introduction}}

\IEEEPARstart{D}{eep} learning technology has made significant progress on various cognitive tasks, but the vast adoption also reveals its shortcomings, such as limited generalizability and lack of interpretability. Witnessing the performance saturation of early models such as multi-layer perceptron (MLP), convolutional neural network (CNN), and recurrent neural network (RNN), one notable recent innovation in deep learning architecture is transformer\cite{vaswani2017attention}. It has two good properties towards artificial general intelligence over conventional models. First, the performance of transformer models continues to grow with their model sizes and training data. Second, transformers can be pre-trained with tons of unlabeled data either through unsupervised or self-supervised learning and can be fine-tuned quickly for each application. With the above characteristics, transformer-based models quickly become mainstream in natural language processing (NLP) as well as other applications such as image classification\cite{dosovitskiy2020image} and object detection\cite{carion2020end}, achieving higher accuracy than other deep-learning models.

Text generation is one of the most popular applications in natural language processing, in which its task is to generate human-readable and plausible sentences from input words called tokens. Text generation is highly demanded in the conversational AI industry, including chatbots\cite{schulman2022chatgpt} and article writing\cite{guardian_2020}. It is also essential for automatic code generation within the context of no-code development paradigm\cite{sanchis2019low}.
Among many transformer models, a decoder-only transformer model from OpenAI named generative pretrained transformer (GPT)\cite{radford2019gpt2, brown2020gpt3} achieves notable performance in text generation. GPT models are pretrained on massive internet data with enormous parameter numbers, raising the quality of text generation close to the human level.


\begin{figure}[!t]
\centering
\includegraphics[width=3.5in]{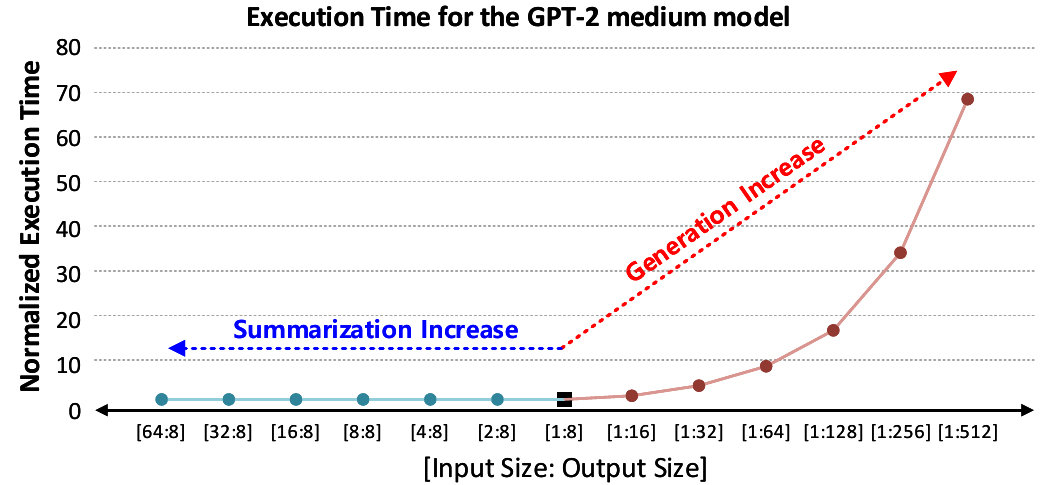}
\caption{The execution time by input and output size for the GPT-2 medium model on a GPU.}
\label{fig_1}
\end{figure}

The text generation process consists of summarization and generation stages. In the summarization stage, all input tokens are computed simultaneously to generate a single output token that will be the input token of the generation stage. On the other hand, the generation stage sequentially generates output tokens through iterations (i.e., an output token per iteration) because the output token becomes the input token of the next iteration.
Figure~\ref{fig_1} shows the execution time of the text generation based on the GPT-2 medium model running on Nvidia Titan RTX GPU, where the input size (i.e., the number of input tokens) and output size (i.e., the number of output tokens) vary. As the graph shows, an increase in the output size linearly increases the total execution time, while an increase in the input size has little impact. This result is because the GPU can efficiently handle a batch of input tokens, but it cannot perform parallel processing in the generation stage due to its sequential nature. In addition, the GPT operations in each iteration of the generation stage are mostly memory-bounded, with a large model size. 
Furthermore, the size of transformer models has a trend of exponential increase (e.g., max 1.5 billion parameters in GPT-2~\cite{radford2019gpt2} and 175 billion parameters in GPT-3~\cite{brown2020gpt3}). It is easily concluded that the performance of text generation depends on the number of iterations in the generation stage and the system's effective memory bandwidth.

Processing-in-memory (PIM) is a promising solution for the memory bandwidth problem. By performing operations near memory, PIM can utilize higher memory bandwidth, hence improving performance significantly for memory-bound operations. Accordingly, many PIM works have been proposed to accelerate memory-bound applications\cite{seshadri2017ambit, gao2019computedram, zhou2022transpim, lee2021hardware, he2020newton, li2017drisa, lenjani2020fulcrum, kim2022overview}. Most studies have focused on accelerating the bit-wise operation and general matrix-vector multiplication (GEMV). It is reasonable because GEMV is a typical memory-bound operation and occupies a large portion of operations. However, the transformer-based models also include additional non-linear functions, which require complex operations that are hard to compute using multiplication and addition. The non-linear functions are compute-bound operations than memory-bound operations, and these functions are not negligible in the entire execution time.

All things considered, there are two ways to achieve higher speedup in PIM. First, the higher bandwidth for memory-bound operation enables higher speedup. GEMV operations for larger weights and biases require higher bandwidth. Second, acceleration for non-linear functions also significantly improves overall performance. Not only do non-linear functions take up a large part, but if PIM accelerates GEMV operation, the proportion of non-linear functions in total execution time increases further. Moreover, supporting non-linear functions in PIM enables accelerating end-to-end model inference. End-to-end acceleration in PIM removes data movement for the intermediate data and prevents the overhead caused by switching between PIM operation and host (generally CPU or GPU) operations\cite{devaux2019true, lee2021hardware}.

The acceleration of transformer-decoder-based generative models in PIM faces several challenges due to its characteristics. First, previous bank-level PIMs have limited bandwidth, and the triple-row activation scheme induces a high latency. It is necessary to use higher bandwidth with low latency to cope with the number of parameters that continue to grow. Second, integrating hardware units in PIM for all necessary operations of the model is challenging. PIM cannot use sufficient metal and area because it is implemented in DRAM technology. Therefore, PIM should support complex operations with limited hardware units. Lastly, data movement is time-consuming, and it is a critical overhead for the entire execution time in PIM. Performing operations simultaneously at the bank-level causes data movement between banks. Furthermore, data movement between PIM and the host also affects overall execution time. Accordingly, supporting end-to-end model inference in PIM can eliminate wasteful data movement.

To address the above challenges, we propose SAL-PIM, a subarray-level PIM architecture for accelerating the end-to-end transformer-decoder-based generative model in PIM. We make the following contributions.

\begin{itemize}[leftmargin=*]
\setlength\itemsep{0.01em}
\item {We present the SAL-PIM architecture that includes subarray-level ALUs (S-ALUs), bank-level units, and channel-level ALUs (C-ALUs). Two types of ALUs (S-ALU and C-ALU) are integrated into the subarray-level and the channel-level, respectively. S-ALU utilizes higher bandwidth than bank-level PIM to compute memory-bound operation, and C-ALU supports accumulation and reduce-sum operation for multiple banks, eliminating data movement between banks.}
\item {We propose acceleration for non-linear functions using a look-up table (LUT)-based linear interpolation in PIM. It enables the computation of complex functions using S-ALUs without additional hardware. Moreover, we optimized a few subarrays, named LUT-embedded subarray, in the bank for LUT operations in DRAM.}
\item {We present a mapping scheme across subarray, bank, and channel for the SAL-PIM architecture. The mapping method enables higher utilization for subarray-level computation and removes data reshaping operations, such as transpose operation, using two input feeding methods and two accumulation directions.}
\item {We evaluated the SAL-PIM architecture using a simulator based on the latest DRAM simulator\cite{kim2015ramulator}. SAL-PIM achieves a maximum 4.72$\times{}$ speedup and an average 1.83$\times{}$ speedup compared to GPU in text generation (input size 32 to 128 and output size 1 to 256) with the GPT-2 medium model. Furthermore, in order to verify the feasibility of PIM, we implemented units of the SAL-PIM architecture in 28-nm logic technology and scaled to 20-nm DRAM technology.}
\end{itemize}
\section{Background}
\label{sec:background}
\subsection{Workload Analysis of GPT-based Text Generation}
\begin{figure}[!t]
\centering
\includegraphics[width=3.5in]{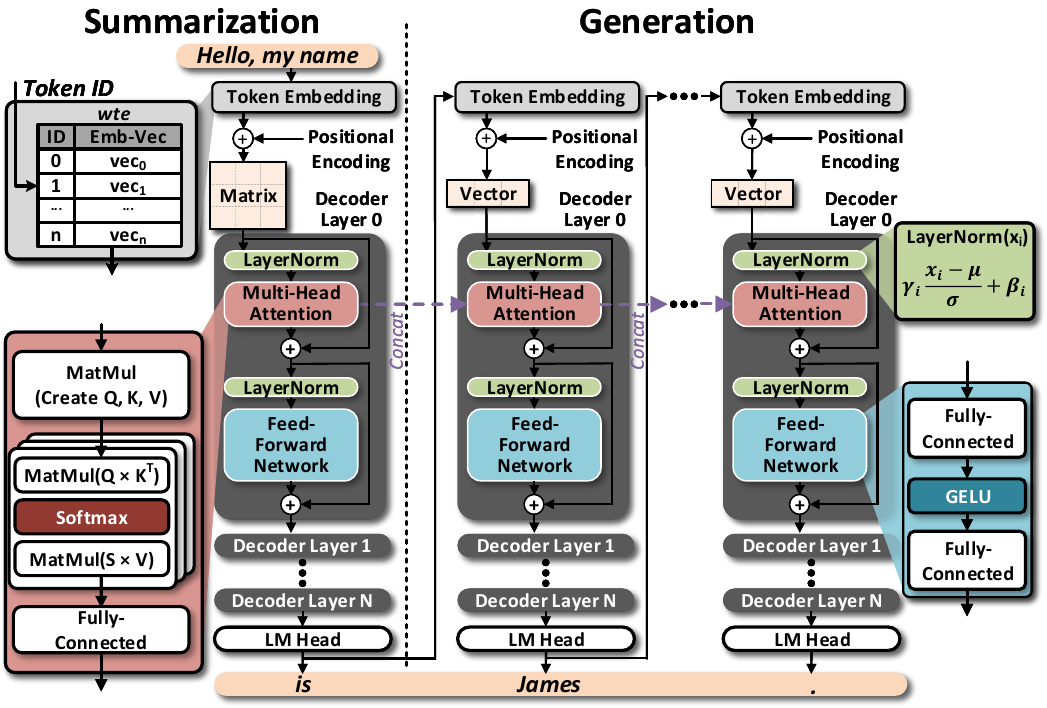}
\caption{GPT structure and text generation process.}
\label{fig_2}
\end{figure}

Text generation's task is generating sentences from given input words. Figure \ref{fig_2} shows the text generation process and structure of GPT. As aforementioned, text generation consists of two stages: summarization and generation. The summarization stage simultaneously computes input tokens (\textit{"Hello, my name"}), as shown in Figure \ref{fig_2}. Each token is embedded in a vector, so input tokens are embedded in a matrix. Therefore, general matrix-matrix multiplications (GEMMs) are mainly performed in the summarization stage, and the operations are compute-bounded due to the parameters being reused for vectors in the embedded matrix. At the end of the summarization stage, GPT makes one output token (\textit{'is'}). Unlike the summarization stage, the generation stage's input is only the former output token (\textit{'is'}). The one input token is embedded into one vector. Thus, the overall operation in the generation stage consists of GEMVs which are memory-bound operations because the input is one vector. Furthermore, the generation stage sequentially generates output tokens, and the model's parameters should be loaded repeatedly for each token generation. Correspondingly, the higher memory bandwidth improves overall inference speed.

GPT is one of the most famous transformer-decoder-based generative models. As shown in Figure \ref{fig_2}, GPT comprises an embedding layer and decoder layers. In the embedding layer, input tokens are converted to vectors through the embedding table and added with positional vectors by the token's position. After passing through the embedding layer, the vector (generation stage) or matrix (summarization stage) goes to the decoder layers. The decoder layer consists of a multi-head attention (MHA), a feed-forward network (FFN), layerNorms, and residual operations. 

\subsubsection{Multi-Head Attention (MHA)}
An MHA is the most time-consuming layer for GPT in text generation. In the MHA, query (\textit{Q}), key (\textit{K}), and value (\textit{V}) are derived from the input vectors. Generated \textit{K} and \textit{V} are concatenated with \textit{Ks} and \textit{Vs}, which are generated in the former token generations, respectively. Accordingly, the number of \textit{K} and \textit{V} increases as token generation progresses. The MHA consists of three steps. First, \textit{Q}, \textit{K}, and \textit{V} are computed from the input vectors through multiplication with weights and addition with biases. The generated \textit{Q}, \textit{K}, and \textit{V} are separated by heads and calculated between each other from the same head. Second, \textit{Q} is multiplied by \textit{K$^T$}, and the result is called the score. The score is masked by the token position and generates an attention score (\textit{S}) through softmax. Then \textit{S} is multiplied by \textit{V}, and the results are concatenated for all heads. \textit{Q}$\times{}$\textit{K$^T$} and \textit{S}$\times{}$\textit{V} are GEMV in each head, but due to the transpose, directions of the matrix-vector multiplication are different. Lastly, the concatenated vectors are computed by a fully-connected layer, then added with residual input.

\subsubsection{Feed-Forward Network (FFN)}
An FFN is composed of two fully-connected layers and an activation function. The fully-connected layers perform multiplication with weight and addition with bias. Its intermediate vector size is four times larger than the input vector of the decoder layer, so the weight and bias are larger than the fully-connected layer of MHA. Moreover, GELU, which consists of complex functions (\textit{tanh} and \textit{sqrt}), is usually used as an activation function. These complex functions are generally done with series calculations, which are time-consuming. Therefore, the FFN occupies a large part of the execution time of the decoder layer due to the large GEMV and complex operations in the activation function.

\begin{figure}[!t]
\centering
\includegraphics[width=3.5in]{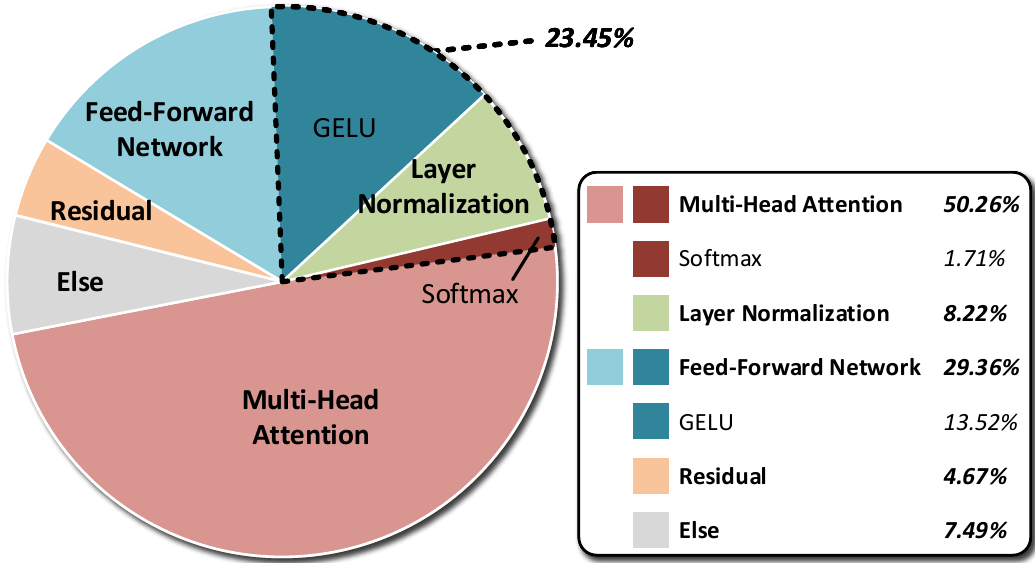}
\caption{The breakdown of execution time for the GPT-2 medium model on a GPU.}
\label{fig_3}
\end{figure}

\subsubsection{Layer Normalization (layerNorm)}
The remained parts of the decoder layer are two layerNorms. The layerNorm is a non-linear function like the softmax in the MHA and GELU in the FFN. Firstly, average and standard deviation are computed position-wise for the input vector. Then, each data of the input vector is subtracted from the average and divided by the standard deviation. The result goes to a fully-connected operation to generate an output vector.\newline 

Figure \ref{fig_3} shows the breakdown of execution time for the GPT-2 medium model on Nvidia Titan RTX GPU. The ratio of the MHA and FFN is 50.26\% and 29.36\%, respectively, which accounts for the most significant portion. Therefore, a higher speedup can be expected when the MHA and FFN are accelerated with higher bandwidth, so previous studies aimed to accelerate GEMV, the main operation of the MHA and FFN, showed notable improvement. However, as shown in Figure \ref{fig_3}, the non-linear functions are also important in the total execution time. Softmax in the MHA, GELU in the FFN, and layerNorm are the non-linear function, which is a compute-bound operation, occupying 23.45\%. In addition, if the MHA and FFN are accelerated, the portion of non-linear functions is further increased. As a result, accelerating the non-linear functions enable higher speedup than only accelerating GEMV.

\begin{table}[!t]
\caption{DRAM-based Processing-in-Memory}
\centering
\includegraphics[width=3.5in]{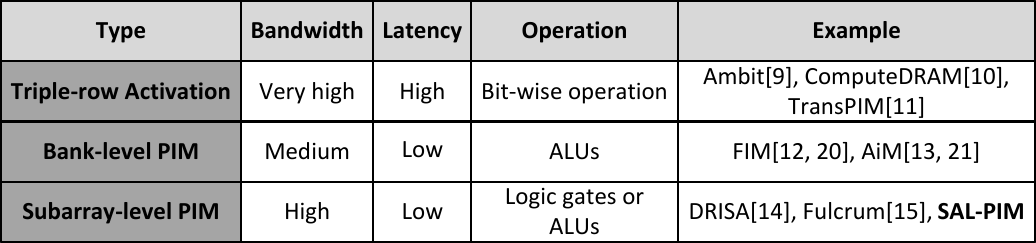}
\label{table_Review}
\end{table}

\subsection{DRAM-based Processing-in-Memory}
Conventional systems using the original DRAM have limited memory bandwidth by a memory standard called JEDEC\cite{standard2013hbm}. In machine learning applications, the limitation remarkably affects the overall latency because most operations are memory-bound operations, such as multiplication and addition, on a tremendous number of data. PIM enables leveraging higher bandwidth thanks to performing computation nearby memory cells. Accordingly, PIM has been actively studied to exploit the higher bandwidth.\cite{seshadri2017ambit, gao2019computedram, zhou2022transpim, lee2021hardware, he2020newton, li2017drisa, lenjani2020fulcrum, kim2022overview}.

Table \ref{table_Review} summarizes DRAM-based PIMs. PIM architecture has two types. The first type computes data using existing memory operations, such as activation, read, and precharge, without additional ALU. For example, Ambit\cite{seshadri2017ambit} and ComputeDRAM\cite{gao2019computedram} perform the bit-wise functions by sharing bit-line with triple-row activation. These accomplish impressive speedup using enormous bandwidth. TransPIM\cite{zhou2022transpim} also adopts triple-row activation with subarray-level adder tree, accelerating transformer models. However, it is challenging to fully utilize bandwidth on complex operations for a small-size vector, and the latency is long because row activation is repeated in DRAM. Moreover, in order to compute complex functions, a Taylor series-like approach\cite{zhou2022transpim} is applied, which requires a lot of additions and multiplications.

The second type computes data using ALUs in the memory. FIM\cite{lee2021hardware, kwon202125} and AiM\cite{he2020newton, lee20221aim} perform MAC operations in the bank, and these are optimized to compute GEMV. However, the number of parameters in the model is increasing, and the higher bandwidth makes PIM expect a high speedup. Therefore, bank-level parallelism is insufficient and should be extended to the subarray-level. There are also previously proposed subarray-level PIM studies\cite{li2017drisa, lenjani2020fulcrum}. DRISA\cite{li2017drisa} implements logic gates or adders in subarray and achieves higher bandwidth. Nevertheless, it is not appropriate for computing non-linear functions. Fulcrum\cite{lenjani2020fulcrum} puts an ALU in a subarray, which is flexible with the operation by subarray. Although Fulcrum has flexibility, which can perform various operations in PIM, it is more critical for PIM to target transformer-based models to optimize memory-bound operations, such as GEMV. As a result, a PIM structure that can perform memory-bound operations with high bandwidth and accelerate non-linear function is required.

\subsection{Linear Interpolation}
\begin{figure}[!t]
\centering
\includegraphics[width=3.5in]{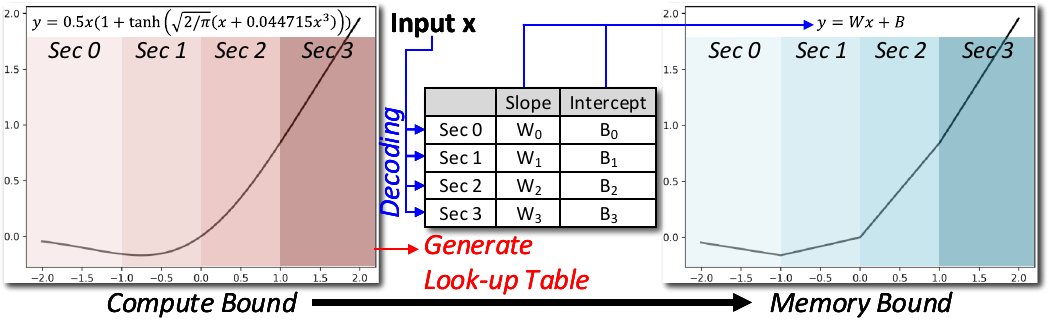}
\caption{Linear interpolation with a look-up table for GELU.}
\label{fig_4}
\end{figure}

Linear interpolation is one of the approximation methods to compute complex functions with limited hardware. Non-linear functions in GPT have complex functions, such as \textit{exp}, \textit{tanh}, and \textit{sqrt}. Hence, to accelerate the non-linear functions in PIM, a method of computing these with limited hardware is required because PIM has limited area and power to implement computing units. Figure \ref{fig_4} shows linear interpolation for GELU. The range of input data is divided into sections, and an LUT stores pre-calculated slopes (W) and intercepts (B) for each section. Then, when the input comes in, find the section that the input belongs to through decoding. The slope and intercept of the corresponding section are multiplied and added with input, respectively. Thus, the complex functions can be calculated with only one multiplication and addition. In other words, the compute-bound operation is projected into the memory-bound operation with LUT, so PIM is appropriate with linear interpolation. Naturally, if the number of sections is insufficient, there is the possibility of accuracy loss. So, we measured accuracy loss by linear interpolation with the GPT-2 medium model on text generation, and the accuracy was kept when the number of sections was larger than 32.

Some previous works use linear interpolation. MVP\cite{junghoon2022MVP} uses dynamic linear interpolation, which dynamically sets the section size, to calculate various types of activation functions. MVP has additional memory to store LUT, but PIM can use DRAM cells as LUT. Also, there is PIM-based linear interpolation. AiM\cite{lee20221aim} applies linear interpolation to the results of adder trees. However, it only computes the activation function. Therefore, to accelerate the end-to-end inference of the model, a PIM structure that can calculate linear interpolation with DRAM cell-based LUT is needed.

\section{SAL-PIM Architecture}
\label{sec:architecture}

SAL-PIM supports three architectural features to address challenges when accelerating the execution of GPT end-to-end. First, it exploits subarray-level parallelism for subarray-level ALUs. The memory-bound operations are computed with a higher bandwidth than the previous bank-level PIM. Second, the SAL-PIM architecture adopts LUT-based linear interpolation, thus minimizing circuit overhead for the complex non-linear functions required. It further eliminates the LUT overhead by reusing the existing DRAM cells. Third, the SAL-PIM architecture includes channel-level ALUs, which perform operations for all banks in the same channel to support the entire GPT operations in memory. Accordingly, data movement between the host and PIM is minimized.

\subsection{Overall Architecture}
\begin{figure}[!t]
\centering
\includegraphics[width=3.5in]{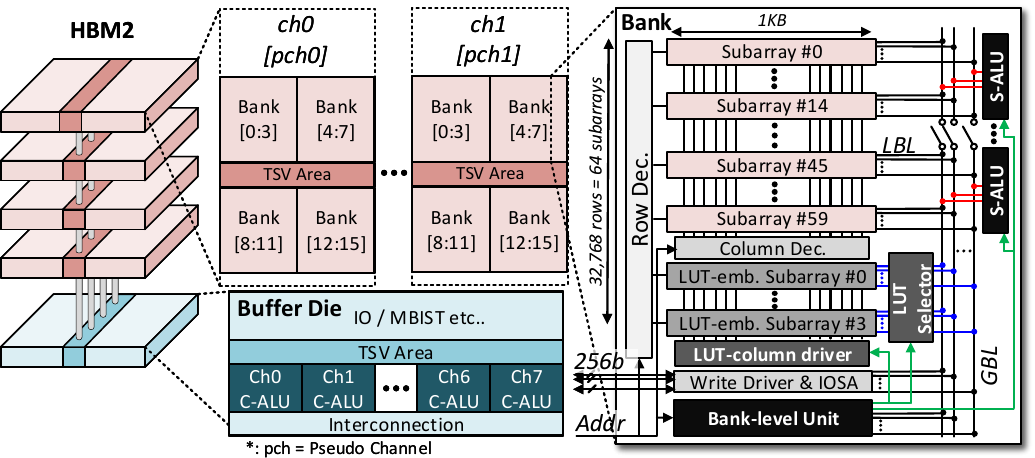}
\caption{Overall architecture of SAL-PIM.}
\label{fig_5}
\end{figure}

The overall architecture of SAL-PIM is based on HBM2, as shown in Figure \ref{fig_5}. The SAL-PIM architecture is composed of four DRAM dies and a buffer die, which are connected through silicon vias (TSVs). The DRAM die has a hierarchical structure of channels, banks, and subarrays. Each channel consists of 16 banks, which are connected to data buses shared with the other banks in the same channel. Hence, the original HBM2 cannot access multiple banks simultaneously. In contrast, the SAL-PIM architecture can access multiple banks simultaneously by integrating computation logic units near each bank. Each bank includes subarrays with 512 rows\cite{kim2012salp}. Each subarray reads data to a bit-line sense amplifier (BLSA) connected to the local bit-lines (LBLs) for the column address. Then, the LBLs are connected to global bit-lines (GBLs), which are used as the path for subarray-level ALUs. In addition, the SAL-PIM architecture includes channel-level ALUs and interconnection, which connects channels, on the buffer die.

The SAL-PIM architecture includes three types of logic units: subarray-level ALU (S-ALU), bank-level unit, and channel-level ALU (C-ALU).
\begin{itemize}[leftmargin=*]
\setlength\itemsep{0.01em}
\item{S-ALU is responsible for principle operations, such as element-wise addition, element-wise multiplication, MAC, and max, with subarray-level parallelism. In order to exploit subarray-level parallelism, the SAL-PIM architecture employs previous research\cite{kim2012salp}, utilizing multiple subarrays. Multiple subarrays in the bank are activated simultaneously by using a BLSA as a cache for each subarray. Ideally, simultaneously operating all of the subarrays can utilize the maximum bandwidth. However, integrating S-ALU into all subarrays causes impractical area overhead, so the subarrays in the bank are grouped to use the S-ALU. For example, if the number of S-ALU is 4 in a bank, the subarray group consists of 15 subarrays without LUT-embedded subarray. The S-ALU is connected to the GBLs to receive weight from memory. Therefore, when multiple subarrays are activated, these are floated and transfer data to each S-ALU, while GBLs generally operate as a single connected data path. Consequently, the SAL-PIM architecture enables utilizing higher internal bandwidth. 

In addition, linear interpolation for the non-linear functions is also performed by multiplication and addition using S-ALU. As aforementioned in Section~\ref{sec:background}, the non-linear functions are compute-bound which occupies a large portion of the total execution time. The SAL-PIM architecture uses LUT-based linear interpolation to reduce the computation intensity and accelerate non-linear functions. In the SAL-PIM architecture, the S-ALU computes linear interpolation by multiplying slope and adding intercept while subarrays are used as LUT. Thus, the acceleration of non-linear functions is achieved in the SAL-PIM architecture without additional hardware.}

\item{A bank-level unit supports feeding input data for S-ALU and generating select signals for LUT-embedded subarray. MHA in GPT needs two types of matrix multiplication (\textit{Q}$\times{}$\textit{K$^{T}$} and \textit{S}$\times{}$\textit{V}), so the SAL-PIM architecture has two directional input feeding schemes for eliminating transpose operation. The bank-level unit selects whether to broadcast the same input for all MACs in S-ALUs or different inputs for each MAC in S-ALUs. Also, decoding units in the bank-level units generate select signals for LUT-embedded subarray. When the LUT-embedded subarrays operate like conventional memory, the decoding units decode addresses. In contrast, when they act as LUT, the decoding units decode data from a register in the bank-level unit.}

\item{C-ALU performs accumulation and reduce-sum operations for multiple banks. To maximize the utilization of S-ALUs, All banks in SAL-PIM operate in parallel. Hence, the SAL-PIM architecture needs to merge each bank's output, so C-ALU merges the outputs of banks and broadcasts the results to all banks in the same channel. C-ALU minimizes data movement between each bank in the same channel.}
\end{itemize}

As a result, by using these three types of units, SAL-PIM supports all operations of the transformer-decoder-based generative model, including non-linear functions, fully-connected operations, and multi-head operations. The detailed circuit of each unit is described in Section \ref{sec:circuit}.
\subsection{Data Mapping}
\label{sec:mapping}

SAL-PIM adopts flexible data mapping schemes across various layers of GPT to run end-to-end inference efficiently. The data mapping of SAL-PIM considers the hardware architecture and dataflows, maximizing the utilization of increased bandwidth. Moreover, the flexible data mapping minimizes the data movement that occurs when the outputs are distributed back to the subarrays for the next layer as inputs. As mentioned in Section~\ref{sec:background}, the decoder layer of GPT consists of FFN, MHA, and layerNorms. These are decomposed into three types of computations (i.e., non-linear function, matrix-vector operation, and multi-head operation), where each has a different dataflow. In addition, the SAL-PIM architecture supports three levels of hierarchy that can operate in parallel (i.e., channels, banks, and subarrays), each of which has different physical characteristics. The higher level hierarchically, the further away from the memory cell, the higher the cost of moving data. For example, data movement between channels is more time-consuming because data moves between them through interconnection on the buffer die. Thus, each channel should be mapped with independent weight, which does not require accumulation. For the banks, the output can be merged in C-ALU, so the weight is mapped to utilize parallelism supremely. The subarrays are tightly coupled, allowing faster and wider data movement among the subarrays. Thus, SAL-PIM can adopt data mapping, which performs the accumulation of partial sums among the subarrays. 


Data mapping schemes of the SAL-PIM architecture reflect three considerations. First, SAL-PIM supports the data mapping scheme that minimizes data movement within and between the computations. Since the weight data of GPT is large and must be tiled, the additional data movement is incurred by partial sums among the tiles, which are generated within each computation. Furthermore, data movement between the computations is incurred since the output of each computation is used as the input of the following computation. Second, SAL-PIM's data mapping also maximizes bandwidth utilization for memory-bound operations. Most operations in the generation stage have no reuse of weight at all, which can be executed faster with higher bandwidth. The extended bandwidth from subarray-level parallelism is fully utilized through data mapping. Third, SAL-PIM supports the data mapping scheme that eliminates any data movement of intermediate data for the two featured operations of MHA, such as concatenation and transposition. Concatenation needs frequent data movement for the concatenated matrices because it is performed by reading all of the data and concatenating them. The data mapping removes the data movement by sequentially mapping them to banks. Also, transposition is time-consuming due to reshaping extensive data and requires additional buffers. The SAL-PIM architecture eliminates transpose operation using data mapping and input feeding schemes.

\begin{figure}[!t]
\centering
\includegraphics[width=3.5in]{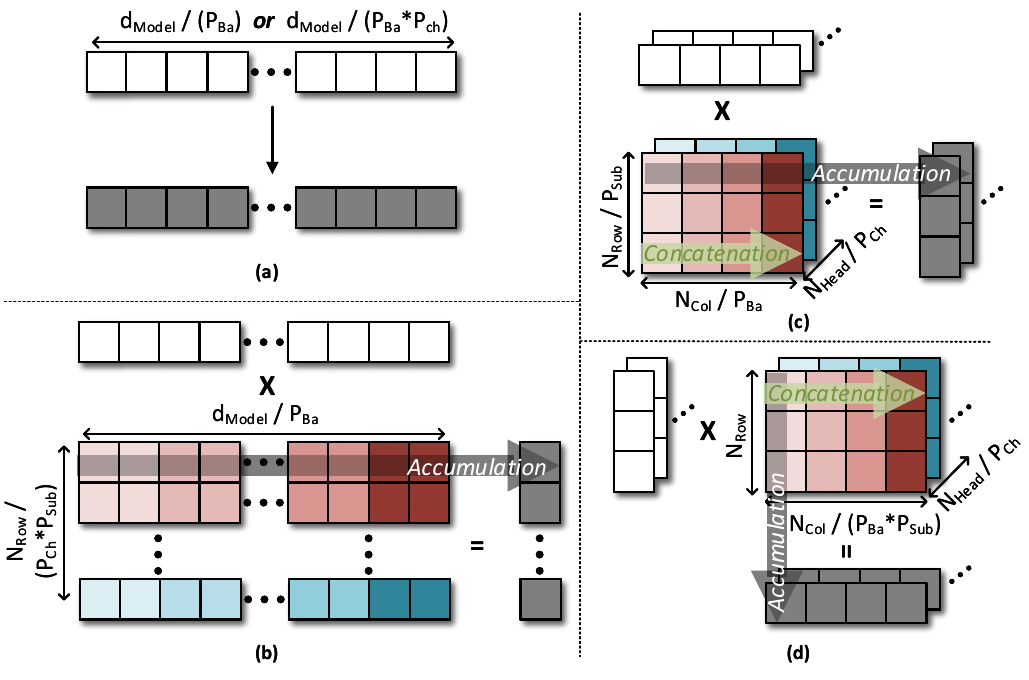}
\caption{Data mapping schemes for SAL-PIM. (a) Non-linear function. (b) Matrix-vector operation. (c) Multi-head operation accumulated by subarrays. (d) Multi-head operation accumulated by banks.}
\label{fig_6}
\end{figure}

Figure \ref{fig_6} illustrates the data mapping schemes for SAL-PIM. The parameters associated with the data mapping in SAL-PIM are $P_{Ch}$, $P_{Ba}$, and $P_{Sub}$, and each refers to parallelism by channels, banks, and S-ALUs, respectively. 

First, the data mapping scheme reduces the data movement within and between computations. As shown in Figure \ref{fig_6}(b), \ref{fig_6}(c), and \ref{fig_6}(d), the SAL-PIM architecture is mapped column of matrix or head of the multi-head operation to $P_{Ch}$, minimizing data movement for partial sums within the computation. Furthermore, data movement caused by the accumulation of partial sums is reduced between banks in the same channel using C-ALU. In addition, the output of computations leads to the input of the following computations in GPT. In particular, the non-linear functions are in the middle of MHA and FFN, so these functions are connected to the other computations directly. As shown in Figure \ref{fig_6}(a), in the case of the non-linear function, two data mapping schemes exist to minimize data movement when the non-linear function continues to other computations. For example, if the matrix-vector operation is followed by the non-linear function, the input vector is tiled in banks and duplicated in channels. On the contrary, if the multi-head operation is followed, the input vector is tiled both banks and channels. The mapping schemes for non-linear functions support the same tiling shapes as the other computations input, so data movements are eliminated. Therefore, the flexible data mapping schemes enable the non-linear function's output to seamlessly be used as input for the following computation.


In addition, the data mapping scheme maximizes bandwidth utilization. In the case of the non-linear function, the number of computations is small due to linear interpolation. So, SAL-PIM cannot utilize the bandwidth fully, and then the data mapping scheme aims to minimize data movement. On the other hand, in the case of matrix-vector and multi-head operations, which are memory-bounded. Therefore, these operations utilize all three types of parallelism, as shown in Figures \ref{fig_6}(b), \ref{fig_6}(c), and \ref{fig_6}(d). For the matrix-vector operation, the rows of the matrix are mapped according to $P_{Ch}$ and $P_{Sub}$, and the columns are divided into $P_{Ba}$. In contrast, for the multi-head operation, the heads, which are independent of other heads, are mapped on $P_{Ch}$, and $P_{Ba}$ and $P_{Sub}$ divide the row or column of the matrix. Both cases need accumulation between banks, and SAL-PIM uses C-ALUs to simplify bank-level accumulation without data movement between banks.

Lastly, the data mapping scheme enables two operations in MHA for text generation: concatenation and transposition. The multi-head operation concatenates \textit{Ks} and \textit{Vs} in text generation in order to project the previous word in the generation of the next word. Therefore, the hardware must enable supporting concatenation of \textit{K} and \textit{V}, and in SAL-PIM, concatenation is performed by mapping the bank sequentially, as shown in Figure \ref{fig_6}(c) and \ref{fig_6}(d). Furthermore, in the multi-head operation, two types of matrix multiplication are needed (\textit{Q}$\times{}$\textit{K$^{T}$} and \textit{S}$\times{}$\textit{V}). To accelerate these, PIM should support transpose operation for matrix, which is time-consuming and requires an additional buffer. In contrast, SAL-PIM supports two accumulation directions and two input feeding methods, which eliminates the need for transpose operations in PIM (Figure \ref{fig_6}(c) and \ref{fig_6}(d)). 

\subsubsection{Dataflow for GPT in SAL-PIM}
These data mapping schemes reduce data movement and enable acceleration of the overall computation of the depicted GPT in Figure 2 using PIM. The input tokens of GPT generate an output token through the embedding layer, positional encoding, layerNorm, MHA, FFN, and residual addition. Firstly, the input token is translated into a vector through the embedding operation. The embedding operation simply finds the vector corresponding to the dictionary, which can be performed by reading the vector at the corresponding address in DRAM. The vector is divided into banks and added with a position vector duplicated in all channels. Then, the vector performs layerNorm. layerNorm subtracts the mean and divides the vector by the standard deviation, followed by the GEMV operation. The mean and standard deviation are obtained through reduction operation in the S-ALU and C-ALU and linear interpolation to compute the reciprocal square root. The mean and standard deviation are broadcasted to all banks and subtracted and multiplied, respectively. The calculated vector performs GEMV operation within its channel, and then the computed vector is broadcasted to all channels for further computations.

Moving on to the MHA, it consists of \textit{Q}, \textit{K}, \textit{V} generation, \textit{Q}$\times{}$\textit{K$^{T}$}, softmax, \textit{S}$\times{}$\textit{V}, and a GEMV operation. Firstly, \textit{Q}, \textit{K}, and \textit{V} generations perform GEMV operation three times each. At this stage, the heads are divided on each channel, allowing independent operations for each head. \textit{Q} is duplicated to all banks, and \textit{K} and \textit{V} values are sequentially mapped to the banks and concatenated. Then, \textit{Q}$\times{}$\textit{K$^{T}$} is performed as shown in Figure 6(d), and the results are stored back in each bank. The computed data is linearly interpolated for exponential calculation and summed in the C-ALU for softmax operation. The sum is broadcasted to all banks, where it performs linear interpolation for the reciprocal operation and is multiplied to compute \textit{S}. Finally, \textit{S} is multiplied by \textit{V} in each bank as depicted in Figure 6(c), and the output is accumulated across all banks in C-ALU. Then, the GEMV operation is performed to complete MHA.

The output of the MHA is reshaped into a single channel, and layerNorm is applied again after residual addition. The resulting vector is then broadcasted across all channels, similar to the previous stage. The FFN stage consists of two GEMV operations and an activation function(GELU). The FFN stage consists of two GEMV operations and an activation function called GELU. The two GEMV operations are performed as shown in Figure 6(b), and the GELU activation is applied using linear interpolation. The resulting vector from the FFN stage undergoes another residual addition. All decoder layers in GPT are configured in the same manner, so SAL-PIM iterates these operations. As a result, SAL-PIM can minimize data movement and maximize parallelism in the GPT inference.
\section{In-Memory Circuit Design}
\label{sec:circuit}
SAL-PIM has three compute units (S-ALUs, Bank-level units, and C-ALUs) and optimized subarrays (LUT-embedded subarrays). This chapter describes detailed circuits and operation flow. 

\begin{figure}[!t]
\centering
\includegraphics[width=3.5in]{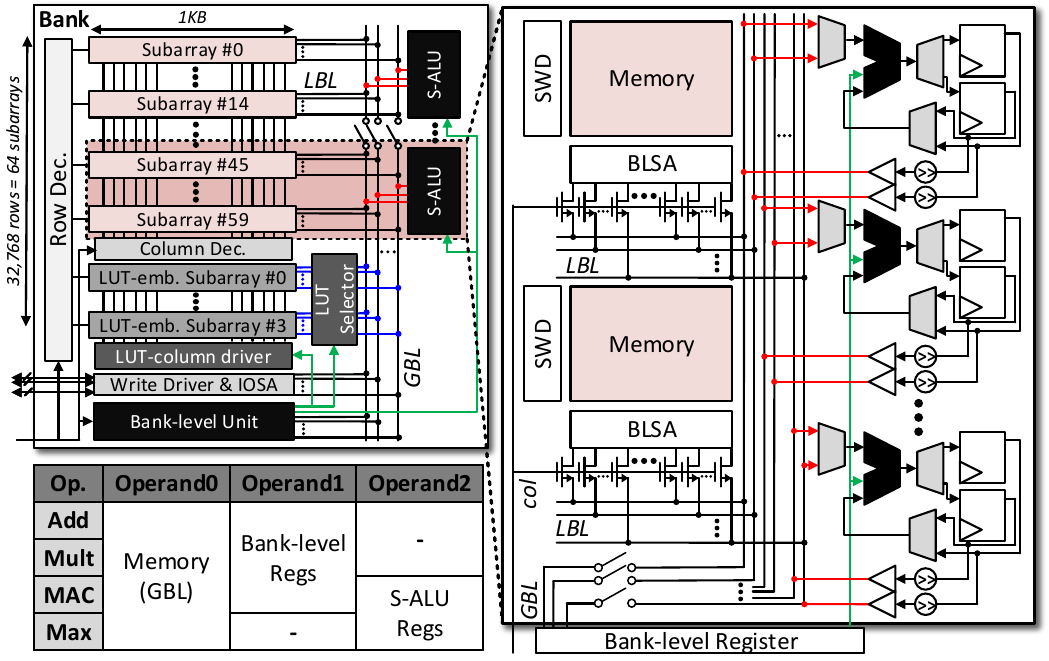}
\caption{Circuit design of S-ALU.}
\label{fig_7}
\end{figure}
\subsection{Subarray-level ALU (S-ALU)}
As aforementioned in Section \ref{sec:architecture}, S-ALU performs principal arithmetic operations of SAL-PIM with subarray-level parallelism. Figure \ref{fig_7} shows the circuit of S-ALU. S-ALU consists of 16-bit fixed-point MAC units, 16$\times{}$32-bit registers, and right shifters. The S-ALU's operation proceeds in three steps. First, read data from memory to the bank-level unit via GBLs, and the data is used as input for S-ALUs. Second, S-ALU calculates the data read from memory and the input broadcasted from the bank-level unit. The results are stored and accumulated in S-ALU's registers. Lastly, the data of the S-ALU's registers are written in the memory. When writing the data from S-ALU's registers, the results stored in the registers are 32-bit precision, but the GBL has half of the precision. Therefore, the results are shifted and truncated by fraction bit using shifters. Then, the tri-state buffer connects the register's output and the GBLs to write the result to the memory.

As shown in the table of Figure \ref{fig_7}, S-ALU's MAC supports four operations: element-wise addition and multiplication, MAC, and max. The max operation is used for linear interpolation of exponential in softmax. The exponential function has a wide range of data; thus, it is computed using linear interpolation after subtracting by the maximum. The MAC units enable the computation of three types of operands from memory (GBL), a bank-level register in the bank-level unit, and registers in S-ALU. The primary operand of S-ALU is data from memory, and GBL is connected during a single operation to receive data from all subarrays. Therefore, one S-ALU is performed at a time to compute linear interpolation using LUT-embedded subarrays. On the contrary, when S-ALUs compute simultaneously in the case of the matrix-vector operation or the multi-head operation, the control signal cuts off the GBLs connected with other S-ALUs. So the S-ALUs operate individually read memory at a time.

In general, most transformer-based models use 32-bit floating-point precision. However, in the previous work\cite{zadeh2020gobo}, the transformer models are sufficient with 8-bit precision. Furthermore, the high bit precision of the register enables minimizing the data loss of accumulation. In order to verify accuracy when fixed-point precision is used, we evaluated the accuracy of the lambada dataset\cite{paperno2016lambada} using 16-bit fixed-point precision. As a result, the accuracy only dropped about 2.8\%  on the GPT-2 medium model without other quantization schemes.

The subarray-level parallelism enables the PIM architecture to achieve outstanding performance, but it has a drawback for the area overhead. As $P_{Sub}$ increases, the area overhead increases proportionally. So, SAL-PIM uses shared MACs in S-ALU, which leverage faster computation units than memory read. Several recent studies, including\cite{wang2021spatten, ham2021elsa}, have utilized MAC units operating at a frequency of 1GHz in logic technology. Moreover, in the recent PIM research\cite{cho2020mcdram} uses 1GHz ALU by utilizing faster transistors in DRAM peripheral area. On the other hand, ALUs in recent PIM works mostly can be only utilized at 250MHz. In HBM2, the clock frequency is 1GHz, and HBM2 reads data on $t_{ccds}=2t_{ck}$ (500MHz) for maximum bandwidth with bank interleaving. However, in PIM, the all-bank mode operates more slowly than the bank-interleaving case because the same banks are consecutively accessed. Accordingly, the memory reads or writes data on $t_{ccdl}=4t_{ck}$ (250MHz), whereas the frequency limitation of ALU is higher than the memory read or write speed. Therefore, by computing part of the data several times while memory moves to the next address, the total number of MACs is reduced, and the area is optimized. For example, 16$\times{}$16-bit inputs are read from memory at one read, and the 16$\times{}$16-bit MACs are required. However, 8$\times{}$16-bit MACs can compute the inputs in two computations thanks to a faster speed (500MHz). We implemented shared MAC using a standard cell library to verify its feasibility, achieving a clock frequency of over 800MHz. Considering the 22\% performance degradation between DRAM and logic technology\cite{kim1999assessing}, it is confirmed that the shared MACs are feasible. As a result, when we implemented the S-ALU, the 8 MACs on 500MHz were about 30\% smaller than the 16 MACs on 250MHz. 

\subsection{LUT-embedded Subarray}
\begin{figure}[!t]
\centering
\includegraphics[width=3.5in]{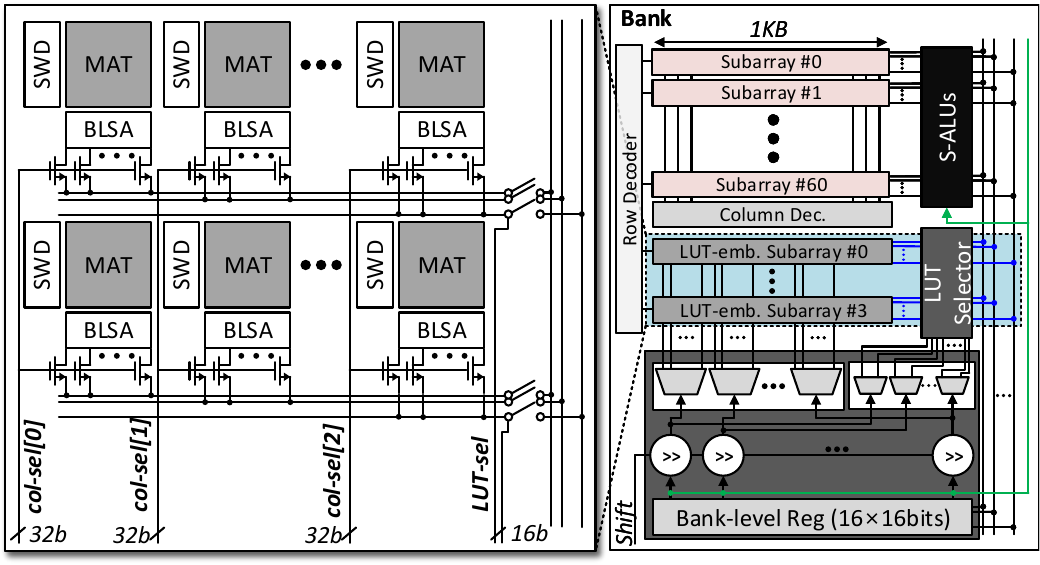}
\caption{Circuit of LUT-embedded subarray and bank-level unit.}
\label{fig_8}
\end{figure}
SAL-PIM uses LUT-based linear interpolation to compute complex functions. However, the original DRAM is unsuitable for performing LUT operations. In the DRAM, a subarray is divided into several MATs\cite{seshadri2019dram}. All MATs of the subarray are operated simultaneously, and a column-select signal selects which column in the MAT is connected to the GBLs. However, LUT needs different addresses for the various data. Suppose the original subarray is used as LUT. In that case, it is necessary to repeatedly approach the subarray as many times as the number of data to find the corresponding value in LUT. These are inefficient tasks for DRAM, and these are time-consuming due to a lot of activation and precharge. Therefore, an LUT-embedded subarray is optimized for LUT operation in DRAM. 

The circuit of the LUT-embedded subarray is shown in Figure \ref{fig_8}. The LUT-embedded subarray is almost identical to the original subarray except for the path to the column-select signal. The LUT-embedded subarray has different column-select signal paths for each MAT, and each MAT receives a different column-select signal. The column-select signals are generated from a column address or data in the bank-level register. However, although MATs operate independently on column addresses, it is not a sufficient solution if one row cannot store all of LUT due to the many sections for linear interpolation. Accordingly, the LUT-embedded subarrays are activated together, and an LUT selector determines which LUT-embedded subarray is connected with GBLs. As a result, a number of LUT-embedded subarrays in the bank and the number of column addresses enable using many sections for linear interpolation.

\begin{figure}[!t]
\centering
\includegraphics[width=3.5in]{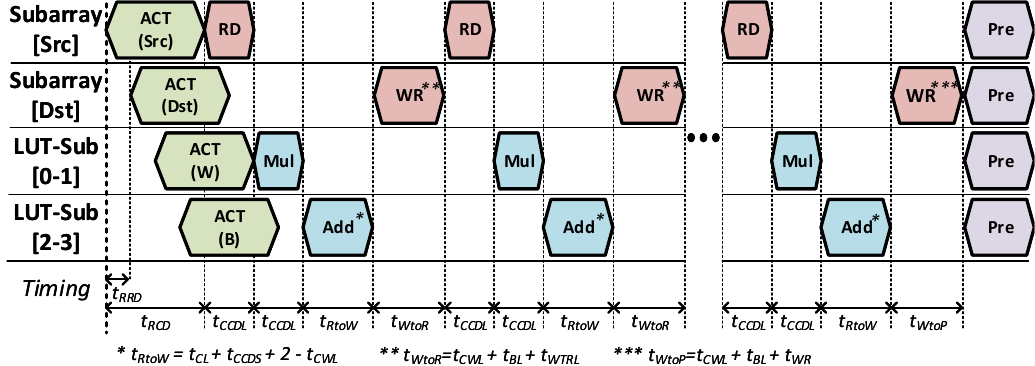}
\caption{Operation flow of LUT-embedded subarray.}
\label{fig_9}
\end{figure}

Figure \ref{fig_9} shows the operation flow of the LUT-embedded subarray. The timing parameter is the same as the parameters in the original DRAM. The operation flow consists of four parts. First, activate all rows for source, destination, slope (W), and intercept (B). Thanks to existing research\cite{kim2012salp}, it is possible to activate simultaneously for different subarrays. Second, read source data from the memory to the bank-level register. The data is used for generating column-select signals and LUT-select signals. Third, S-ALU multiplies the slopes and adds intercepts for linear interpolation. The data read from the LUT-embedded subarray moves through GBL and is computed in an S-ALU. Lastly, the data of register in S-ALU, results of linear interpolation, is written to memory cells. At the end of the operation, all subarrays are precharged. As shown in Figure \ref{fig_9}, the activation and precharge commands are issued once at the start and the end, respectively. Consequently, the LUT-embedded subarray enables linear interpolation for the entire data in bank-level register at a time, which can be up to 16 times faster than the original DRAM, which performs linear interpolation done one by one.

\subsection{Bank-level Unit}
The bank-level unit is responsible for generating select signals to perform LUT-embedded subarrays and broadcasting input data to S-ALUs. The bank-level unit consists of a bank-level register and decoding units. The bank-level register broadcasts the input to S-ALUs in the same bank with two methods. The first is each data of the bank-level register goes to each MAC of S-ALU. It is for element-wise computations. The other method is only single data of the bank-level register is broadcasted to all of the MACs in the bank. It is used in MAC operation and enables each MAC to accumulate the output.

Furthermore, data in the bank-level register is used as addresses of LUT-embedded subarray. Figure \ref{fig_8} also shows the circuit of the bank-level unit. The decoding units in the bank-level unit have two types: column decoder and sub-sel decoder. Each is responsible for generating column-select signals and LUT-select signals, respectively. Making the select signals through decoding in an appropriate bit position for fixed-point bit precision is possible. The bit position means linear interpolation region. If the slope and intercept are generated in intervals from -4 to 4, the decoding units decode addresses based on the interval. Therefore, the right shifters select the bit position since each function's proper linear interpolation range differs.

\begin{figure}[!t]
\centering
\includegraphics[width=3.5in]{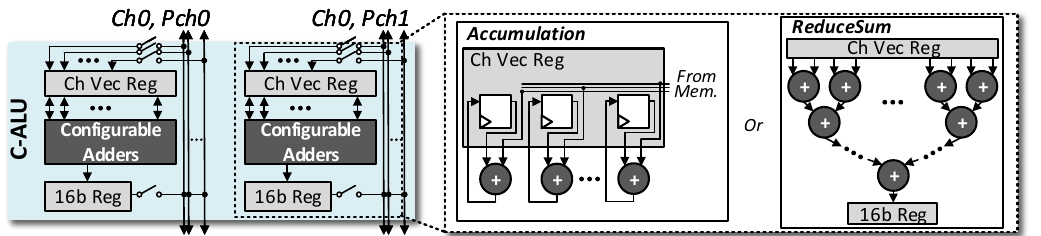}
\caption{Circuit design of channel-level ALU.}
\label{fig_10}
\end{figure}

\subsection{Channel-level ALU (C-ALU)}
C-ALU supports merging operations for multiple banks in the same channel. Because the SAL-PIM architecture operates multiple banks simultaneously, the merging procedures for banks should be performed. When the data is moved from PIM to the host, the energy caused by data movement is consumed, and PIM should wait for the host's work. In addition, since the summed result is used as input for the following operation, the result returns immediately from the host to the PIM and is broadcasted to the banks. These operations, such as the accumulation of banks and reduce-sum operation, are performed in C-ALU, which is implemented in the buffer die of HBM2 for each channel. At the buffer die, although there is no bandwidth, it is possible to eliminate data movement by performing simple addition. Likewise, if bank-level adder trees exist, PIM can reduce the execution time for merging operations. However, since accumulations by banks are small in GEMV, it is wasteful to implement it at the bank, considering the area overhead.

Figure \ref{fig_10} illustrates the circuit of C-ALU, which consists of two channel vector registers (16$\times{}$16-bit), two scalar registers (16-bit), and configurable adders. The configurable adders have sixteen adders, which act as two types of modules, either an accumulator or an adder tree, depending on the command. In the case of accumulation, the configurable adders accumulate memory output to the channel vector register. The accumulated data is broadcasted to memory or summed into one data. In the case of reduce-sum operation, the configurable adders operate as an adder tree to compute summation about the channel vector register. The summed result is stored in the channel scalar register. The channel scalar register's data is broadcasted to memory for layerNorm or softmax. 
\section{Evaluation}
\label{sec:evaluation}

\subsection{Methodology \& Configuration}
We evaluated the SAL-PIM architecture on the simulator modified from the latest DRAM simulator\cite{kim2015ramulator}. The SAL-PIM is based on conventional HBM2\cite{standard2013hbm}, and the configuration with timing parameters\cite{o2017fine} is shown in Table \ref{table_1}. As shown in Table \ref{table_1}, we evaluated the SAL-PIM architecture of $P_{Sub}$=1, 2, and 4, so the maximum bandwidth is four times larger than bank-level PIM when $P_{Sub}$ is 4. In addition, the SAL-PIM architecture applied linear interpolation with 64 sections on GELU, \textit{exp}, \textit{sqrt}, and reciprocal operations. As aforementioned, when the number of sections is larger than 32, it has no accuracy drop by linear interpolation.

The transformer-decoder-based generative model we used for the evaluation is the GPT-2 medium model (with 345 million parameters)\cite{radford2019gpt2}, which computes the vector with a size of 1,024 and has 24 decoder layers. The model's overall structure is the same as in Figure \ref{fig_2} of Section \ref{sec:background}. We compared the SAL-PIM architecture to the server-level GPU, Nvidia Titan RTX\cite{TitanRTX}. The GPU used 24GB GDDR6 memory with 1.77GHz clock frequency. The maximum available bandwidth is 672GB/s, 2.63$\times{}$ larger than the maximum bandwidth of HBM2. The GPU executed the GPT-2 medium model from FasterTransformer Framework\cite{FasterTransformer} for comparison baseline. 

\begin{table}[!t]
\caption{SAL-PIM configuration}
\centering
\includegraphics[width=3.5in]{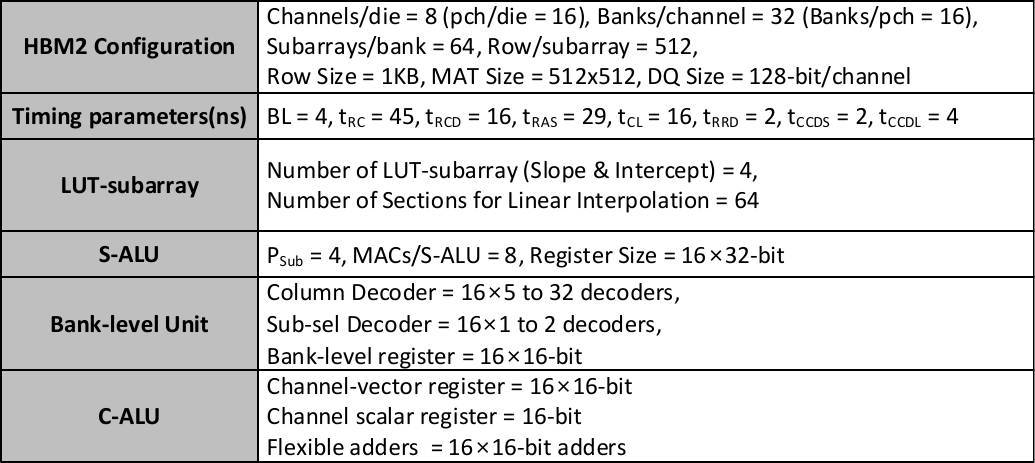}
\label{table_1}
\end{table}


\begin{figure*}[!t]
\centering
\includegraphics[width=7.16in]{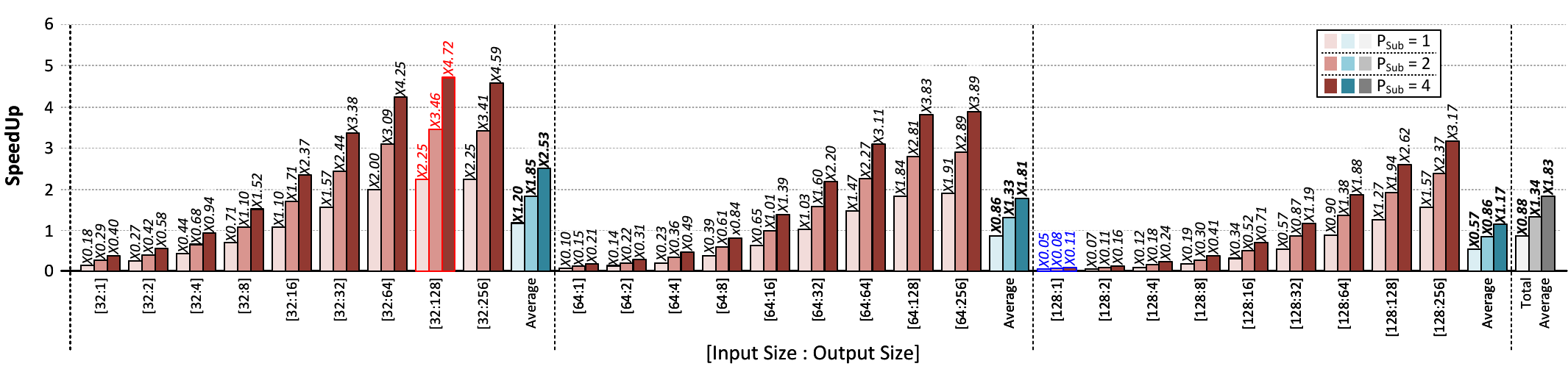}
\caption{Speedup of SAL-PIM compared to GPU for text generation by input and output sizes.}
\label{fig_11}
\end{figure*}

\subsection{Area \& Power}
Firstly, we implemented units of the SAL-PIM architecture to verify its feasibility. These units are implemented using standard cells of the TSMC 28-nm technology with Design Compiler. Since the previous HBM2-PIM\cite{kwon202125} was fabricated on the Samsung 20-nm DRAM technology, we scaled the implemented 28-nm area to the 20-nm area on DRAM technology. Prior research\cite{kim1999assessing, li2017drisa, lenjani2020fulcrum, zhou2022transpim} has shown that the area difference between DRAM technology and CMOS logic technology is approximately 1.8 times. To be conservative, we further increased this area overhead by 3.6 times, doubling the initial scaling factor. Additionally, we measured the power consumption of the units using the PrimePower tool\cite{PrimePower}.

Table \ref{table_2} summarizes the area and power consumption of each unit. Accordingly, When the SAL-PIM has $P_{Sub}$=4, the number of S-ALU is 64 in a channel, and the area overhead is 4.81\% compared to conventional HBM2\cite{zhou2022transpim}. The overhead is far below than 25\% threshold mentioned in previous work\cite{he2020newton}. Accordingly, the area overhead is acceptable, and SAL-PIM does not need to sacrifice memory capacity. Furthermore, we checked the power consumption of SAL-PIM architecture. When the power consumption is maximum, all of S-ALU performs MAC operations simultaneously, it is only 9.04\% of the total power budget of HBM2\cite{o2017fine}. The results indicate that the SAL-PIM architecture is feasible with the permissible area and power overheads.

\begin{table}[!t]
\caption{Summarization of area and power consumption\\of units in the SAL-PIM architecture}
\centering
\includegraphics[width=3.5in]{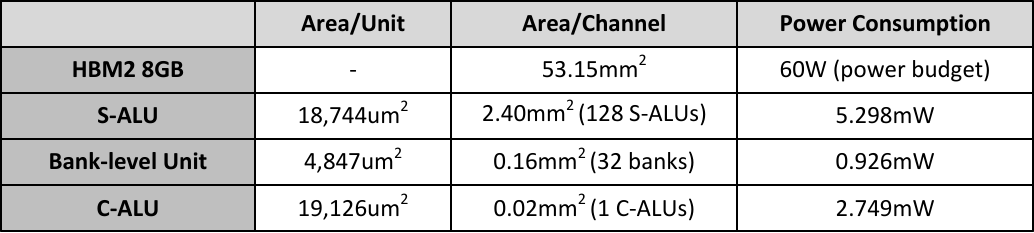}
\label{table_2}
\end{table}

\subsection{Comparison with GPU}
We evaluated the performance of the SAL-PIM architecture for text generation by various input and output sizes. As aforementioned in Section~\ref{sec:background}, when the input size is large, the number of the input vectors in the summarization stage is increased, and when the output size is large, the number of token generations in the generation stage is increased. Figure \ref{fig_11} shows the speedup compared to the GPU by input and output sizes. The input lengths are 32, 64, and 128 tokens, and the output length is from 1 to 256 tokens. The input and output sizes were determined by the typical ranges of user requests based on previous work\cite{hong2022dfx}. In addition, because the SAL-PIM architecture supports GPT end-to-end, the overhead from data movement is considered on the simulator. Therefore, comparing the SAL-PIM architecture using the software simulator with GPU is reliable.

As shown in Figure \ref{fig_11}, in the case of increasing the input size, The speedup decreases. The decrement in performance is because the increase in summarization means an increase in operations that can be performed in parallel, and the high-performance GPU performs operations much faster. However, SAL-PIM operates on a slower frequency, and the number of ALU is far less than the GPU. 

On the contrary, in the case of increasing the output size, the speedup tends to grow for the same input size. The improvement is because the time to execute the model once for one input vector is much shorter than that of the GPU. Hence, the longer the generation stage, the higher the speedup of the SAL-PIM. As a result, when $P_{Sub}$ is 4, the maximum speedup of SAL-PIM is 4.72$\times{}$ compared to the GPU when the input size is 32, and the output size is 128. Then, the average speedup is 1.83$\times{}$ compared to the GPU. This result indicates the maximum speedup lower than the gain obtained from subarray-level parallelism. This means means that even though SAL-PIM supports memory-bound operation of generation stage much faster than the GPU, the summarization stage and compute-bound operations degrade overall performance. However, this performance increase in the generation stage is meaningful because the output tokens tend to be much longer in general workloads of text generation application.


\begin{figure}[!t]
\centering
\includegraphics[width=3.5in]{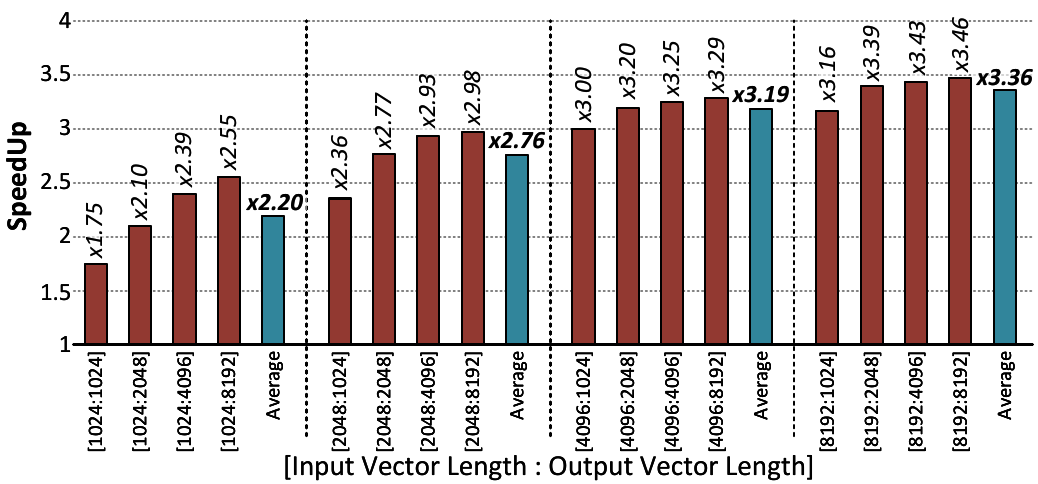}
\caption{Speedup of SAL-PIM for GEMV compared to bank-level PIM.}
\label{fig_12}
\end{figure}

\subsection{Comparison with Bank-level PIM}
SAL-PIM architecture has a much higher maximum bandwidth compared to Bank-level PIM, so we evaluated the SAL-PIM architecture compared to the bank-level PIM architecture to verify whether this high bandwidth is being fully utilized is necessary. The bank-level PIM architecture is based on Newton\cite{he2020newton}, which has multipliers and adder trees in each bank. We modified the SAL-PIM simulator to the Bank-level PIM simulator with the same configuration as Table \ref{table_1}. Figure \ref{fig_12} shows the SAL-PIM's speedup for GEMV operation compared to the bank-level PIM. 

The maximum bandwidth of SAL-PIM is 4$\times{}$ larger than bank-level PIM because of $P_{Sub}$=4. However, in the case of the small size of the input and output vector, the minimum speedup is only 1.75$\times{}$, as shown in Figure \ref{fig_12}. The degradation is because accumulation is needed for the SAL-PIM by mapping scheme, but bank-level PIM does not require the bank-level data movement. Hence, with larger input and output vector sizes, the portion of data movement is smaller, and the speedup is closer to maximum gain. In the GPT-2 medium model, the vector length is only 1,024, but the latest transformer-decoder-based generative model\cite{brown2020gpt3} has a longer vector length of up to 12,288. Therefore, acceleration through subarray-level parallelism is required for a higher performance increase for the large-size model.

\section{Additional Analysis}
\label{sec:additional_analysis}

\subsection{LUT-embedded subarray}
\begin{figure}[!t]
\centering
\includegraphics[width=3.5in]{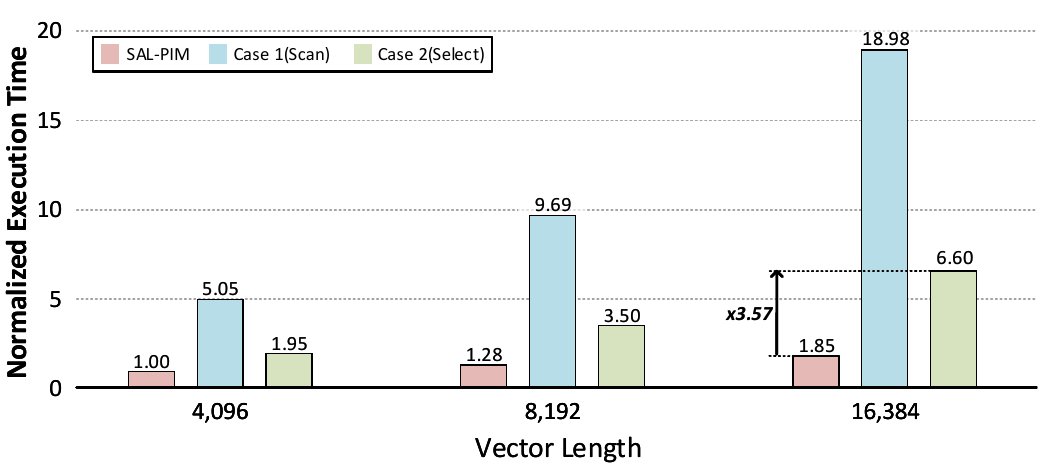}
\caption{Execution time comparison for the LUT operation between LUT-embedded subarray and two other linear interpolation cases.}
\label{fig_13}
\end{figure}

The SAL-PIM architecture uses the linear interpolation method to compute the non-linear functions. Furthermore, SAL-PIM has the LUT-embedded subarray to optimize LUT operations on DRAM. However, the LUT operation can be performed through other methods in the original DRAM subarrays. So. we evaluated the LUT-embedded subarray between the other two cases. Case 1 is \textit{Scan}, which scans all regions of LUT storing slope and intercept. For example, when the number of sections is 64, the slopes and intercepts are stored in 128 addresses. In that case, starting from the beginning, read to the end and find the slope and intercept corresponding to the section. Case 2 is \textit{Select}. This is the case that the LUT decodes the corresponding address from the first data to the last data to which the LUT is applied and finds the slopes and intercepts sequentially. 

Figure \ref{fig_13} shows the Execution time comparison of the LUT-embedded subarray and two linear interpolation cases without the LUT-embedded subarray. Case 1 shows the worst execution time because the size of the bank-level register is limited. Furthermore, in this case, the number of sections is only 64. For a large number of sections, Case 1 is worsened due to the large region to scan. In Case 2, the LUT operation is performed on only one data at a time in a bank. Therefore, In the case of large vectors, more speedup can be expected for the LUT-embedded subarray. Accordingly, the LUT-embedded subarray shows a 3.57$\times{}$ speedup in a case with a vector size of 16,384, as shown in Figure \ref{fig_13}.

\subsection{Subarray-level Parallelism}

\begin{figure}[!t]
\centering
\includegraphics[width=3.5in]{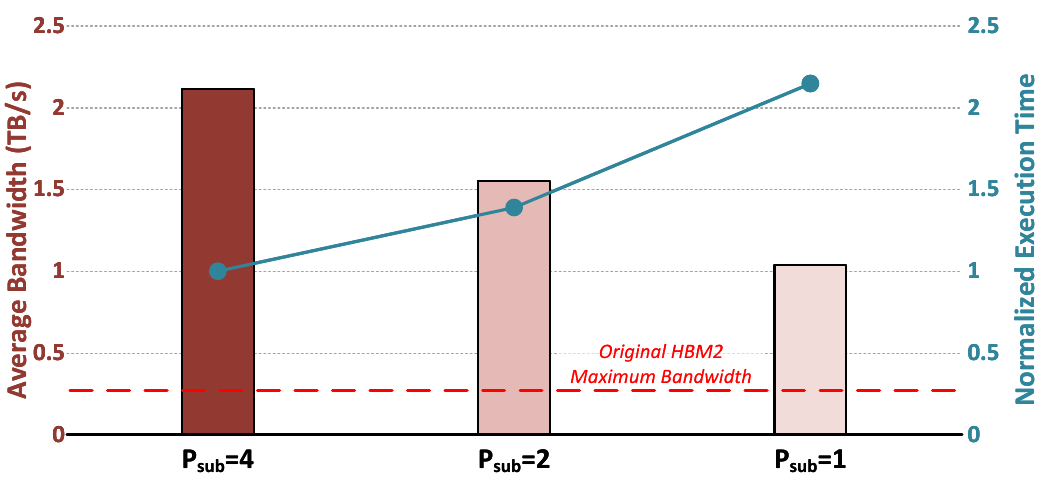}
\caption{The execution time and average bandwidth by subarray-level parallelism on text generation.}
\label{fig_14}
\end{figure}

The subarray-level parallelism enables to use of an enormous bandwidth maximum of 8TB/s when $P_{Sub}$ is 4. However, if the higher bandwidth is not fully used, there is no advantage for the area and power overheads caused by S-ALUs. Therefore, we evaluated the SAL-PIM architecture on various $P_{Sub}$. Figure \ref{fig_14} shows the execution time and average bandwidth by the number of subarray-level parallelisms on text generation. 

The SAL-PIM architecture uses subarray-level parallelism for matrix-vector and multi-head operations, occupying about 60\% of the total execution time in text generation applications. Hence, the SAL-PIM architecture achieves a 2.11$\times{}$ speedup when $P_{Sub}$ is 4 compared to when $P_{Sub}$ is 1, as shown in Figure \ref{fig_14}, showing utilizes higher bandwidth efficiently for memory-bound operations. Similarly, the average bandwidth of the case of $P_{Sub}$ is 4 is also about two times larger than the case of $P_{Sub}$ is 1. Furthermore, considering there are larger models than the GPT-2 medium model, the higher bandwidth can be used efficiently to accelerate memory-bound operations.

\begin{figure}[!t]
\centering
\includegraphics[width=3.5in]{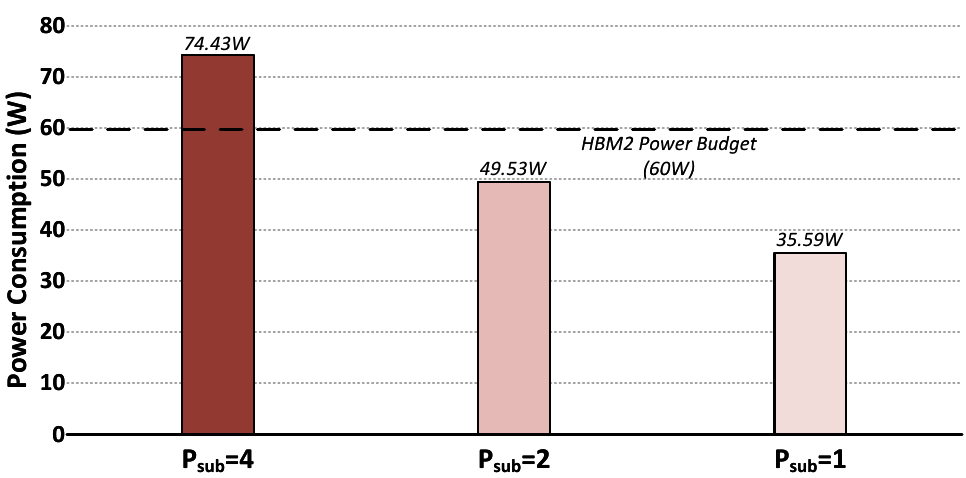}
\caption{The power consumption for subarray-level parallelism in 32 token generations of the GPT-2 medium model.}
\label{fig_review}
\end{figure}

Subarray-level parallelism offers increased bandwidth but consumes more energy. Operating multiple banks and subarrays simultaneously result in higher energy consumption compared to conventional DRAM operations. However, it provides advantages for energy due to the shorter data movement distances. Figure \ref{fig_review} illustrates the power consumption of SAL-PIM for various values of $P_{Sub}$.

In our evaluation, we conducted 32 token generations using the GPT-2 medium model. The energy consumption was assigned as follows\cite{o2017fine}: $e_{act}=909pJ$, $e_{pre-gsa}=1.51pJ/bit$, $e_{post-gsa}=1.17pJ/bit$, and $e_{io}=0.80pJ/bit$. We assumed that 26\% of the total HBM poewr budget is allocated for refresh operations\cite{shin2017extreme}. As depicted in Figure \ref{fig_review}, when $P_{Sub}$ is 1 or 2, the power consumption remains significantly lower than the power budget, while for $P_{Sub}$ is 4, it exceeds the power budget by 24.0\%. Despite this, considering the high power usage expected for computations in CPUs or GPUs, SAL-PIM can still be considered sufficiently power-efficient for transformer-decoder-based text generation. Furthermore, This result assumes that the ALUs are always operating. Therefore, there is room for further power savings through increasing bulk capacitance and optimization techniques such as clock gating or power gating.

\subsection{Future Work}
While SAL-PIM presents a novel architecture for efficient end-to-end text generation within the PIM paradigm, leveraging its inherent high bandwidth, further investigation is necessary to address two key challenges. These areas represent valuable future directions for optimizing SAL-PIM's performance and scalability.

The first challenge concerns the performance bottleneck by the summarization stage. The compute-bound nature of summarization significantly impedes SAL-PIM's overall speedup, diminishing the potential gains from its high bandwidth advantage. To mitigate this, future research should explore mapping and operation strategies that offload the summarization stage to dedicated accelerators like GPUs or NPUs while reserving PIM for the generation stage. This heterogeneous execution scheme could potentially unlock the full performance potential of SAL-PIM.

The second challenge stems from the ever-growing size of modern LLMs, which often exceed the capacity of single PIM units. Consequently, SAL-PIM requires a new approach to accommodate these expansive models. Two promising avenues for future research lie in exploiting parallelism: Intra-PIM parallelism through pipelined weight write, read, and computation operations and Inter-PIM parallelism via synchronization and workload distribution across multiple PIM instances. Additionally, recent research\cite{alizadeh2023llm} has proposed leveraging the inherent bias in output tokens to selectively load weights, minimizing memory footprint. Therefore, sparsity-aware data mapping and handling is another intriguing direction for PIM research.

By addressing these challenges and actively pursuing the outlined future work, SAL-PIM holds immense potential as a groundbreaking architecture for efficient and scalable end-to-end text generation within the PIM domain.

\section{Related Works}
\label{sec:related}

\subsection{Accelerators for Transformer models}
There are previous works for accelerating transformer models. SpAtten\cite{wang2021spatten} accelerates the attention through sparsity handling using cascade token and head pruning. Also, ELSA\cite{ham2021elsa} suggests the architecture with hardware-software co-design for self-attention. These achieve higher performance improvements than GPU but only accelerate attention in the model. DFX\cite{hong2022dfx} proposes the architecture for accelerating end-to-end text generation on multi-FPGA. It was actually implemented and verified using FPGAs and showed a higher performance increase in the generation stage compared to GPU. However, DFX uses many HBMs in FPGAs, so SAL-PIM is a promising solution that uses high bandwidth with fewer HBMs.

\subsection{PIM using Look-up-table}
There are a few architectures PIM using LUT. Lacc\cite{deng2019lacc}, pluto\cite{ferreira2022pluto}, and LT-PIM\cite{zhou2022ltpim} use DRAM cells as LUT and suggest novel circuit designs for LUT operation. Accordingly, these show notable performance for massive computations. However, there is a challenge to calculate using only the LUT. In order to scan cases in LUT, the latency increases exponentially in the case of higher bit-precision. In addition, most machine learning uses a higher bit precision of 16-bit or more. Therefore, SAL-PIM uses LUT-based linear interpolation to support higher bit precision. AiM\cite{lee20221aim} also utilizes LUT-based linear interpolation to support various activation functions. However, it is limited to supporting activation functions and is only applied to the result data of a bank-level adder tree. On the other hand, SAL-PIM supports a wide range of non-linear functions and utilizes LUT-embedded subarrays to apply LUT-based linear interpolation to multiple data at once.
\section{Conclusion}
\label{sec:conclusion}
With the emergence of the transformer, the performance of machine learning has increased faster in several applications. However, as larger pretrained transformer-based models have been developed, the overall execution time has slowed. To address this issue, We propose the SAL-PIM architecture, the first PIM architecture to accelerate the end-to-end transformer-decoder-based generative model. SAL-PIM uses S-ALU to accelerate memory-bound operation by utilizing subarray-level parallelism with the optimized mapping scheme. Moreover, it optimizes the area overhead of S-ALU by sharing MAC units leveraging the slow clock frequency of commands for the same bank. C-ALU is integrated on the buffer die in HBM2 and performs accumulation and reduce-sum operations for multiple banks. Furthermore, SAL-PIM adopts linear interpolation to compute complex non-linear functions to support end-to-end inference. Then, a LUT-embedded subarray is suggested to optimize LUT operation in DRAM.

We have implemented the proposed logic units of SAL-PIM architecture in a 28-nm technology to verify its feasibility. Also, we built the SAL-PIM simulator based on Ramulator to evaluate the architecture. As a result, when the input size is from 32 to 128 and the output size is from 1 to 256, SAL-PIM achieves a maximum of 4.72$\times{}$ speedup and an average of 1.83$\times{}$ speedup for the text generation based on the GPT-2 medium model compared to the GPU.

\bibliographystyle{IEEEtran}
\bibliography{IEEEabrv, references}

\begin{IEEEbiography}
[{\includegraphics[width=1in,height=1.25in,clip,keepaspectratio]{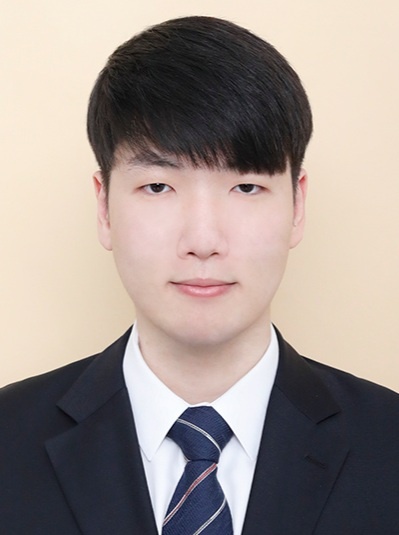}}]{Wontak Han} (Graduate Student Member, IEEE)
received the B.S. degree in Electronic Engineering from Hanyang University, Seoul, South Korea, in 2021, and the M.S. degree in electrical engineering from Korea Advanced Institute of Science and Technology (KAIST), Daejeon, South Korea, in 2023. He is currently an engineer in Samsung Electronics DRAM Design Team.

His research interests include energy-efficient processing-in/near-memory architecture, and deep-learning network and database management system (DBMS) accelerators.
\end{IEEEbiography}

\begin{IEEEbiography}
[{\includegraphics[width=1in,height=1.25in,clip,keepaspectratio]{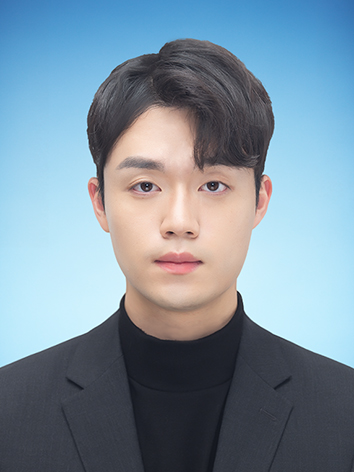}}]{Hyunjun Cho} (Graduate Student Member, IEEE)
received the B.S. degree in Electrical Engineering from Korea Advanced Institute of Science and Technology (KAIST), Daejeon, South Korea, in 2023, where he is currently pursuing the M.S. degree. 

His research interests include energy-efficient processing-in/near memory architecture, design of neural processing units (NPUs), and the design of hardware architectures that preserve security.
\end{IEEEbiography}

\begin{IEEEbiography}
[{\includegraphics[width=1in,height=1.25in,clip,keepaspectratio]{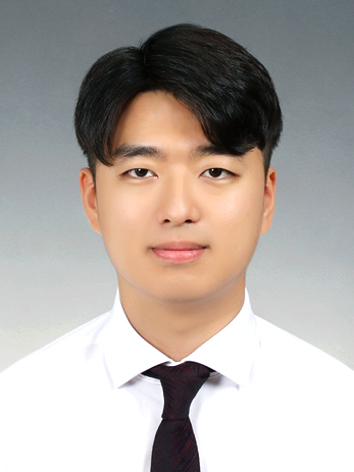}}]{Donghyuk Kim}
(Graduate Student Member, IEEE) received the B.S. degree in electrical and computer engineering from University of Washington, Seattle, USA, in 2020. He received the M.S. degree in electrical and computer engineering from Korea Advanced Institute of Science and Technology (KAIST), Daejeon, South Korea, in 2022, where he is currently pursuing the Ph.D. degree.

His research interests include energy-efficient processing-in-memory architecture for machine learning and database, and deep neural network accelerators.
\end{IEEEbiography}

\begin{IEEEbiography}
[{\includegraphics[width=1in,height=1.25in,clip,keepaspectratio]{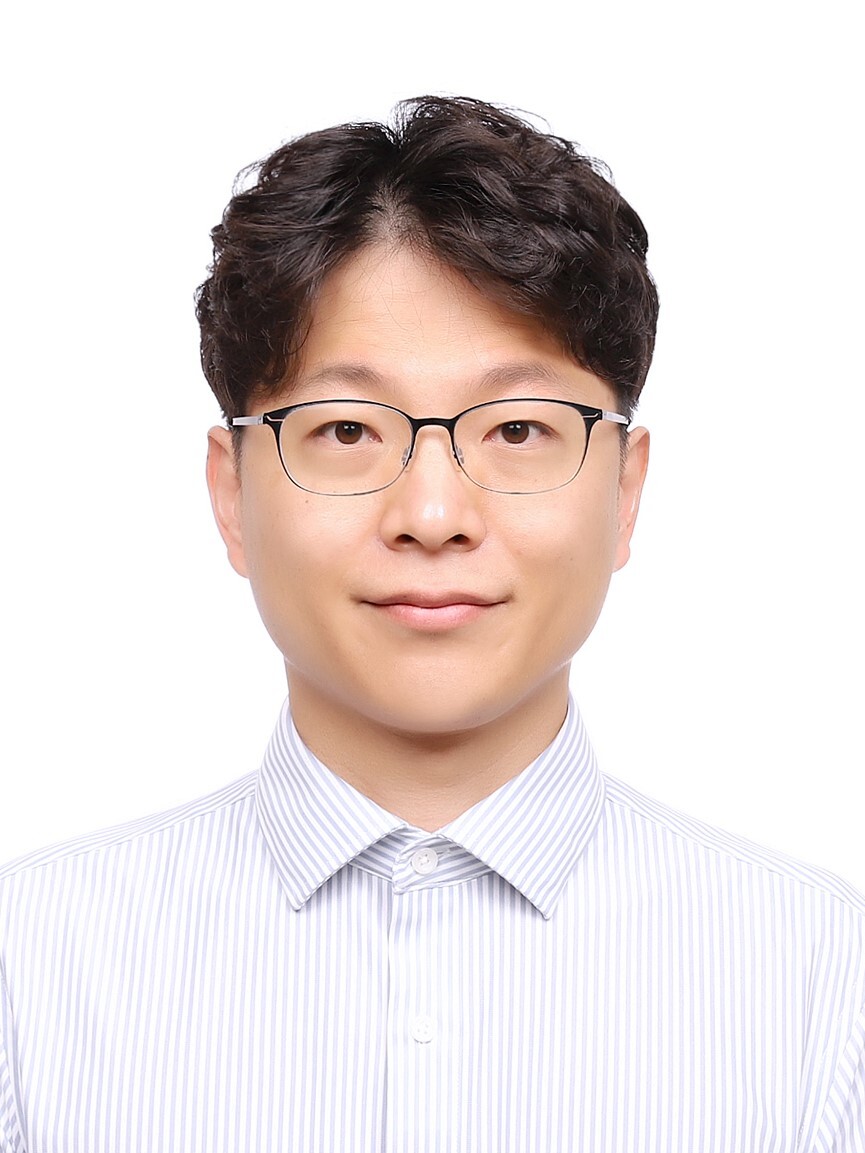}}]{Joo-Young Kim} (Senior Member, IEEE) received the B.S., M.S., and Ph. D. degrees in Electrical Engineering from KAIST, Daejeon, South Korea, in 2005, 2007, and 2010, respectively. He is currently an Associate Professor in the School of Electrical Engineering at KAIST. He is also the Director of AI Semiconductor Systems Research Center. His research interests span various aspects of hardware design, including VLSI design, computer architecture, FPGA, domain-specific accelerators, hardware/software co-design, and agile hardware development. Before joining KAIST, he was a Senior Hardware Engineering Lead at Microsoft Azure, Redmond, WA, USA, working on hardware acceleration for the hyper-scale big data analytics platform named Azure Data Lake. He was also one of the initial members of Catapult project at Microsoft Research, Redmond, WA, USA, where he deployed a fabric of FPGA accelerators in datacenters to accelerate critical cloud services, such as machine learning, data storage, and networking.

\end{IEEEbiography}


\end{document}